\documentstyle[twoside,fleqn,espcrc2,epsf]{article}

\makeatletter
\def\lesssim{\mathrel{\mathpalette\vereq<}}
\def\vereq#1#2{\lower3pt\vbox{\baselineskip1.5pt \lineskip1.5pt
\ialign{$\m@th#1\hfill##\hfil$\crcr#2\crcr\sim\crcr}}}
\def\gtrsim{\mathrel{\mathpalette\vereq>}}
\def\alt{\lesssim}
\def\agt{\gtrsim}
\makeatother

\begin{document}
\thispagestyle{empty}

\title{
\vbox to0pt
{\vskip0pt minus100cm\centerline{\fbox{\parbox{\textwidth}
{\small\noindent To be published in NEUTRINO 98, 
Proc.\ XVIII International Conference on Neutrino Physics and
Astrophysics, Takayama, Japan, 4--9 June 1998,
ed.\ by Y.~Suzuki and Y.~Totsuka.}}}
\vskip2.2cm}\vskip-22pt\noindent
Axion Hunting at the Turn of the Millenium}

\author{Georg~Raffelt\\
        {\ }\\
        Max-Planck-Institut f\"ur Physik, F\"ohringer Ring 6, 
        80805 M\"unchen, Germany}

%\date{\today}

\begin{abstract}
  The status of several current and proposed experiments to search for
  galactic dark-matter and solar axions is reviewed in the light of
  astrophysical and cosmological limits on the Peccei-Quinn scale.
\end{abstract}

\maketitle

%%%%%%%%%%%%%%%%%%%%%%%%%%%%%%%%%%%%%%%%%%%%%%%%%%%%%%%%%%%%%%%%%%%%%%
%% Section I %%%%%%%%%%%%%%%%%%%%%%%%%%%%%%%%%%%%%%%%%%%%%%%%%%%%%%%%%
%%%%%%%%%%%%%%%%%%%%%%%%%%%%%%%%%%%%%%%%%%%%%%%%%%%%%%%%%%%%%%%%%%%%%%

\begin{figure}[b]
\hbox to\hsize{\hss\epsfxsize=7cm\epsfbox{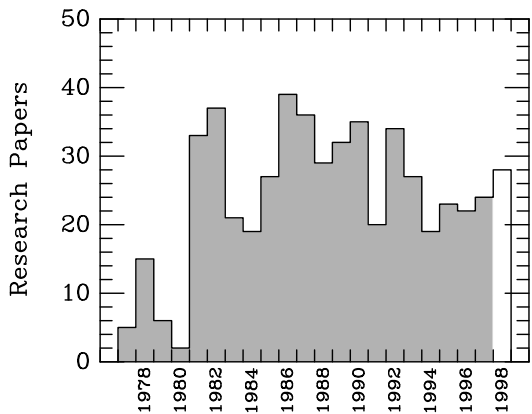}\hss}
\caption{\label{fig:papers} Original research papers in the SPIRES HEP
  data base at SLAC~\protect\cite{SLAC} with {\it axion}, {\it axino},
  or {\it Peccei-Quinn} in their title. The date is when a preprint
  was recorded, not the year of publication in a journal.  The papers
  of 1977 include those of Ref.~\protect\cite{PecceiQuinn} which do
  not use our keywords, and the ones for 1998 are only up to the end
  of June.}
\end{figure}

\section{INTRODUCTION}

Twenty years after their inception~\cite{PecceiQuinn},
axions~\cite{Axions} remain a popular solution to the strong CP
problem as well as a candidate for the cold dark matter of the
universe. As a research topic, axions and related issues seem to have
retained much of their appeal (Fig.~\ref{fig:papers}), and indeed 1998
could yet become the year with the largest number of axion research
papers ever. All that is missing in this flurry of activities is the
appearance of the main character, the axion itself, which thus far has
eluded all attempts at a discovery. 

In the framework of ``invisible axion'' models where the scale $f_a$
at which the Peccei-Quinn symmetry is spontaneously broken could be
arbitrarily large, and where axions therefore could be arbitrarily
weakly interacting, previous experiments really did not stand a
plausible chance of finding these particles.  Therefore, the most
exciting recent development is that there are now two full-scale
search experiments for galactic dark-matter axions in operation which
do have a realistic discovery potential. Also, there is a surprising
amount of activity around the search for solar axions which is
beginning to become competitive with astrophysical limits. The chance
of an actual discovery, however, appears more remote than in the
search for galactic axions.

%%%%%%%%%%%%%%%%%%%%%%%%%%%%%%%%%%%%%%%%%%%%%%%%%%%%%%%%%%%%%%%%%%%%%%
%% Section II %%%%%%%%%%%%%%%%%%%%%%%%%%%%%%%%%%%%%%%%%%%%%%%%%%%%%%%%
%%%%%%%%%%%%%%%%%%%%%%%%%%%%%%%%%%%%%%%%%%%%%%%%%%%%%%%%%%%%%%%%%%%%%%

\section{ASTROPHYSICAL LIMITS}

\begin{figure}[b]
\hbox to \hsize{\hfil\epsfxsize=\hsize\epsfbox{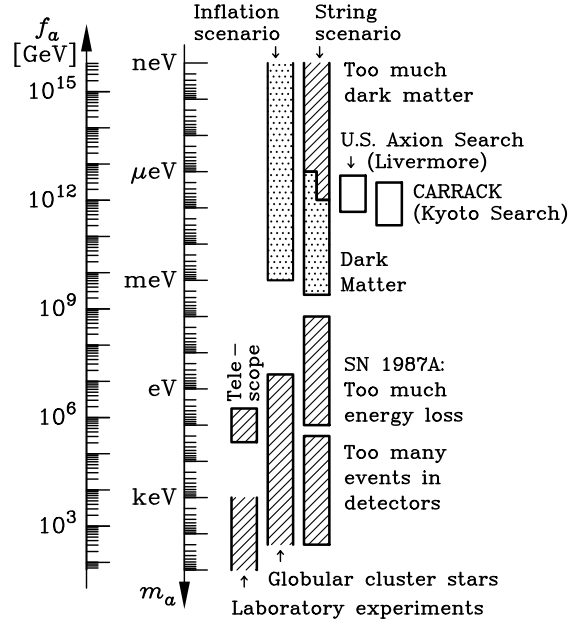}\hfil}
\caption{\label{fig:limits} Astrophysical and cosmological exclusion
regions (hatched) for the axion mass $m_a$ or equivalently, the
Peccei-Quinn scale $f_a$. An ``open end'' of an exclusion bar means
that it represents a rough estimate; its exact location has not been
established or it depends on detailed model assumptions.  The globular
cluster limit depends on the axion-photon coupling; it was assumed
that $E/N=8/3$ as in GUT models or the DFSZ model.  The SN 1987A
limits depend on the axion-nucleon couplings; the shown case
corresponds to the KSVZ model and approximately to the DFSZ model.
The dotted ``inclusion regions'' indicate where axions could plausibly
be the cosmic dark matter.  Most of the allowed range in the inflation
scenario requires fine-tuned initial conditions.  In the string
scenario the plausible dark-matter range is controversial as indicated
by the step in the low-mass end of the ``inclusion bar.''}
\end{figure}

Axion models are characterized by the Peccei-Quinn scale $f_a$, or
equivalently by the axion mass $m_a=0.60~{\rm eV}~(10^7~{\rm
GeV}/f_a)$. Several astrophysical lower limits on $f_a$
(Fig.~\ref{fig:limits}) are based on the requirement that the axionic
energy loss of stars, notably globular-cluster stars or the core of
supernova (SN) 1987A, is not in conflict with certain observed
properties of these objects~\cite{RaffeltBook,MiniReview}.  These
limits imply $m_a\alt10^{-2}~{\rm eV}$ or $f_a\agt10^9~{\rm GeV}$,
indicating that axions, if they exist, are both extremely light and
very weakly interacting.

These limits on the axion mass are indirectly derived from limits on
the coupling strength to photons (globular cluster stars) and nucleons
(SN~1987A). The axionic two-photon interaction is ${\cal
L}_{\rm int}=g_{a\gamma}{\bf E}\cdot{\bf B}\,a$, where
\begin{equation}\label{eq:coupling}
g_{a\gamma}=-\frac{3\alpha}{8\pi f_a}\,\xi=
-\frac{m_a/{\rm eV}}{0.69\times10^{10}~{\rm GeV}}\,\xi
\end{equation}
with
\begin{equation}\label{eq:photoncoupling}
\xi\equiv\frac{4}{3}\left(\frac{E}{N}-1.92\pm0.08\right).
\end{equation}
$E/N$ is a model-dependent ratio of small integers. In the DFSZ
model or GUT models one has $E/N=8/3$, corresponding to $\xi\approx1$,
and it is this case for which the globular-cluster limit 
\begin{equation}\label{eq:globlimit}
g_{a\gamma}\alt 0.6\times10^{-10}~{\rm GeV}^{-1}
\end{equation}
is shown in Fig.~\ref{fig:limits} as a limit on the axion mass,
$m_a\alt0.4~{\rm eV}$.  The
axion-photon coupling for a variety of models has recently been
compiled~\cite{Kim}; often-discussed cases are $E/N=8/3$ (DFSZ) or
$E/N=0$ (KSVZ).

However, models with $E/N=2$ can be constructed, allowing for a near
or complete cancellation of $g_{a\gamma}$.  In this case there is no
globular-cluster limit on $m_a$ or $f_a$ so that there is a small
window for $m_a$ near $10~{\rm eV}$. It was recently
shown~\cite{Moroi} that in this range axions could be a cosmological
hot dark matter component which certain structure-formation arguments
suggest in addition to the main cold dark matter.

In the early universe, axions are thermalized if
$f_a\alt 10^8\,{\rm GeV}$~\cite{Turner87}, a region excluded by the
stellar-evolution limits except for the special case $E/N=2$.  If
inflation occurred after the Peccei-Quinn symmetry breaking or if
$T_{\rm reheat}<f_a$, the ``misalignment
mechanism''~\cite{Misalignment} leads to a contribution to the cosmic
critical density of $\Omega_a h^2\approx 1.9\times 3^{\pm1}
\,(1\,\mu{\rm eV}/m_a)^{1.175}\, \Theta_{\rm i}^2 F(\Theta_{\rm i})$
where $h$ is the Hubble constant in units of $100\,\rm
km\,s^{-1}\,Mpc^{-1}$. The function $F(\Theta)$ with
$F(0)=1$ and $F(\pi)=\infty$ accounts for anharmonic corrections to
the axion potential. Because the initial misalignment angle
$\Theta_{\rm i}$ can be very small or very close to $\pi$, there is no
real prediction for the mass of dark-matter axions even though one
would expect $\Theta_{\rm i}^2 F(\Theta_{\rm i})\sim1$.

A possible fine-tuning of $\Theta_{\rm i}$ is limited by
inflation-induced quantum fluctuations which in turn lead to
temperature fluctuations of the cosmic microwave
background~\cite{Turner91,BS98}. In a broad class of inflationary
models one thus finds an upper limit to $m_a$ where axions could be
the dark matter.  According to the most recent discussion \cite{BS98}
it is about $10^{-3}~\rm eV$ (Fig.~\ref{fig:limits}).

If inflation did not occur at all or if it occurred before the
Peccei-Quinn symmetry breaking with $T_{\rm reheat}>f_a$, cosmic axion
strings form by the Kibble mechanism~\cite{Davis}.  Their motion is
damped primarily by axion emission rather than gravitational waves.
After axions acquire a mass at the QCD phase transition they quickly
become nonrelativistic and thus form a cold dark matter component. The
axion density such produced is similar to that from the misalignment
mechanism for $\Theta_{\rm i}={\cal O}(1)$, but in detail the
calculations are difficult and somewhat controversial between two
groups of authors~\cite{BattyeShellard,SikivieHagmann}.  Taking into
account the uncertainty in various cosmological parameters one arrives
at a plausible range for dark-matter axions as indicated in
Fig.~\ref{fig:limits}.

%%%%%%%%%%%%%%%%%%%%%%%%%%%%%%%%%%%%%%%%%%%%%%%%%%%%%%%%%%%%%%%%%%%%%%
%% Section III %%%%%%%%%%%%%%%%%%%%%%%%%%%%%%%%%%%%%%%%%%%%%%%%%%%%%%%
%%%%%%%%%%%%%%%%%%%%%%%%%%%%%%%%%%%%%%%%%%%%%%%%%%%%%%%%%%%%%%%%%%%%%%

\section{DARK MATTER SEARCH}

\begin{figure}[b]
\hbox to \hsize{\hfil\epsfxsize=3cm\epsfbox{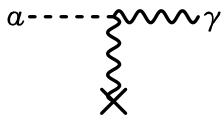}\hfil}
\caption{\label{fig:primakoff} Primakoff conversion of axions into
photons in an external electromagnetic field.}
\end{figure}

If axions are the galactic dark matter one can search for them in the
laboratory. The detection principle is analogous to the Primakoff
process for neutral pions, i.e.\ the two-photon vertex allows for
axion transitions into photons in the presence of an external
electromagnetic field (Fig.~\ref{fig:primakoff}).  Dark matter axions
would have a mass in the $\rm \mu eV$ to meV range. As they are bound
to the galaxy their velocity dispersion is of order the galactic
virial velocity of around $10^{-3} c$ so that their kinetic energy is
extremely small relative to their rest mass. Noting that a frequency
of $1\,\rm GHz$ corresponds to $4\,\rm \mu eV$, the Primakoff
conversion produces microwaves. Galactic axions are nonrelativistic
while the resulting photons are massless so that the conversion
involves a huge momentum mismatch which can be overcome by looking for
the appearance of excitations of a microwave cavity rather than for
free photons.

An axion search experiment thus consists of a high-$Q$ microwave
resonator placed in a strong external magnetic field (``axion
haloscope'' \cite{Sikivie}).  The microwave power output of such a
detector on resonance is~\cite{Sikivie,Krauss}
\begin{eqnarray}
P&\approx&0.4\times10^{-22}\,{\rm Watts}
\left(\frac{V}{0.2\,\rm m^3}\right)\nonumber\\
&\times&\left(\frac{B}{7.7~\rm Tesla}\right)^2
\left(\frac{C}{0.65}\right)
\left(\frac{Q}{10^5}\right)\nonumber\\
&\times&
\left(\frac{\rho_a}{300~\rm MeV~cm^{-3}}\right)
\left(\frac{m_a}{1~\rm \mu eV}\right),
\end{eqnarray}
where $V$ is the cavity volume, $B$ the applied magnetic field, $C$ a
mode-dependent form factor which is largest for the fundamental
T$_{010}$ mode, $Q$ the loaded quality factor, and $\rho_a$ the local
galactic axion density.  If $m_a$ were known it would be easy to
detect galactic axions with this method---one may verify or reject a
tentative signal by varying, for example, the applied magnetic field
strength. Therefore, it would be hard to mistake a background signal
for dark-matter axions.  The problem is, of course, that $m_a$ is not
known so that one needs a tunable cavity, stepping its resonance
through as large a frequency range as possible and to look for the
appearance of microwave power beyond thermal and amplifier noise.

\begin{figure}[b]
\hbox to \hsize{\hfil\epsfxsize=\hsize\epsfbox{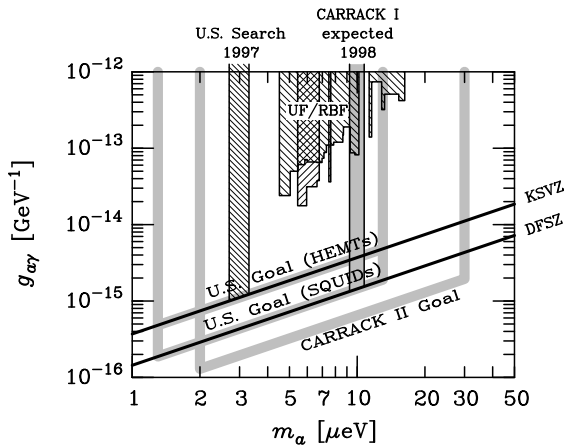}\hfil}
\caption{\label{fig:darkmatter} Limits on galactic dark matter axions
  from the University of Florida (UF) \protect\cite{UFexperiment} and
  the Rochester-Brookhaven-Fermilab (RBF) \protect\cite{RBFexperiment}
  pilot experiments and the recent limit from the U.S.~Axion Search
  \protect\cite{Livermore}. Also shown are the search goals for the
  U.S.\ experiment employing HEMTs for microwave detection, for a next
  generation experiment using SQUIDs, the 1998 search goal for
  CARRACK~I (Kyoto) and for CARRACK~II, both using Rydberg atoms.}
\end{figure}

Two pilot experiments of this sort~\cite{UFexperiment,RBFexperiment}
have excluded the range of axion masses and coupling strengths
indicated in Fig.~\ref{fig:darkmatter}.  For a standard local halo
density of about $300~\rm MeV~cm^{-3}$ they were not sensitive enough
to reach realistic axion models. Two current experiments with
larger cavities, however, have the requisite sensitivity.

The U.S.~Axion Search~\cite{Livermore} uses conventional microwave
amplifiers (HEMTs) which limit the useful cavity temperature to about
$1.4~\rm K$. A first exclusion slice has been
reported~\cite{Livermore}---see Fig.~\ref{fig:darkmatter} where the
ultimate search goal is also shown. In a next-generation experiment
one would use SQUID amplifiers, increasing the sensitivity to
encompass more weakly coupled axion models.

The Kyoto experiment CARRACK~\cite{Kyoto}, on the other hand, uses a
completely novel detection technique, based on the excitation of a
beam of Rydberg atoms which passes through the cavity. This is
essentially a counting method for microwaves which does not require a
(noisy) amplifier so that one can go to much lower physical cavity
temperatures. This enhances the sensitivity and also allows one to use
smaller cavity volumes and thus to search for larger axion
masses. With the current setup a narrow slice of axion masses is to be
searched (Fig.~\ref{fig:darkmatter}), while a new apparatus currently
under construction will allow for the coverage of a much broader mass
range.

The search goals of these second-generation experiments covers the
lower range of plausible axion masses in the framework of the
cosmological string-scenario of primordial axion production, and a
significant portion of the plausible mass range in the inflation
scenario if one does not wish to appeal to fine-tuned initial
conditions of the axion field (Fig.~\ref{fig:limits}). If these
experiments fail to turn up axions, it would be extremely important to
extend the experimental search into a regime of larger masses toward
the meV scale. This would require new detection methods.

\newpage

%%%%%%%%%%%%%%%%%%%%%%%%%%%%%%%%%%%%%%%%%%%%%%%%%%%%%%%%%%%%%%%%%%%%%%
%% Section IV %%%%%%%%%%%%%%%%%%%%%%%%%%%%%%%%%%%%%%%%%%%%%%%%%%%%%%%%
%%%%%%%%%%%%%%%%%%%%%%%%%%%%%%%%%%%%%%%%%%%%%%%%%%%%%%%%%%%%%%%%%%%%%%

\section{SOLAR AXIONS}

\subsection{Helioscope Method}

\vfil

Another classic way to search for axions is to use the Sun as a source
and to attempt an experimental detection of this flux. Unfortunately,
the experimental sensitivity typically lies in an $f_a$ range which is
already excluded by the stellar-evolution limits of
Fig.~\ref{fig:limits} so that one needs to appeal to large systematic
uncertainties of the astrophysical bounds in order to hope for a
positive detection. On the other hand, such experiments can provide
independent limits on the parameters of axions and similar particles
even if the chances for a positive detection seem slim.

In the so-called ``helioscope'' method~\cite{Sikivie,Bibber} one again
uses the Primakoff effect (Fig.~\ref{fig:primakoff}) by pointing a
long and strong dipole magnet toward the Sun. The axions produced in
the hot interior of the Sun would have typical energies of a few keV
and would thus convert into x-rays which can then be picked up by a
detector at the down-stream end of the magnet. A pioneering experiment
was conducted several years ago~\cite{Lazarus}, but detecting axions
would have required a flux larger than what is compatibel with the
solar age.

Recently, first results were reported from the Tokyo axion helioscope
where a dipole magnet was gimballed like a telescope so that it could
follow the Sun and thus reach a much larger exposure
time~\cite{Tokyo}. The limit on the axion-photon coupling of
$g_{a\gamma}\alt 6\times10^{-10}~{\rm GeV}^{-1}$ is less restrictive
than the globular-cluster limit of Eq.~(\ref{eq:globlimit}), but more
restrictive than the solar-age limit of $25\times10^{-10}~{\rm
GeV}^{-1}$~\cite{Dearborn87}, and also more restrictive than a
recent solar limit of about $10\times10^{-10}~{\rm GeV}^{-1}$ which is
based on helioseismological sound-speed profiles of the
Sun~\cite{Schlattl}.

Another helioscope project with a gimballed dipole magnet was begun in
Novosibirsk several years ago~\cite{Vorobev}, but its current status
has not been reported for some time.  A very intruiging project at
CERN would use a decommissioned LHC test magnet that could be mounted
on a turning platform to achieve reasonably long times of alignment
with the Sun~\cite{LHC}.  With this setup one would begin to compete
with the globular cluster limit of Eq.~(\ref{eq:globlimit}).

The helioscope approach is bedevilled by the same problem which
requires the use of a resonant cavity in the galactic axion search,
viz.\ the momentum mismatch between (massive) axions and (massless)
photons in the Primakoff process.  For example, the above limit of the
Tokyo helioscope applies only for $m_a\alt0.03~{\rm eV}$, implying
that the ``axion-line''---the relationship between $g_{a\gamma}$ and
$m_a$ of Eq.~(\ref{eq:coupling})---is not even touched, i.e.\ the
limit applies only to particles which for a given $g_{a\gamma}$ have a
smaller mass than true axions.

In a next step one will fill the transition region with a pressurized
gas, giving the photon a dispersive mass in order to overcome the
momentum mismatch~\cite{Bibber}.  As in the cavity experiments, this
is a resonant method (the match is only good for a small range of
axion masses) so that one needs to take many runs with varying gas
pressure to cover a broad $m_a$ range. In this way it is hoped to
eventually cut across the axion line. The same approach would have to
be used for the proposed CERN helioscope.

\subsection{Bragg Diffraction}

\vfil

An alternative method to overcome the mo\-mentum-mismatch problem in
the Primakoff process is to use an inhomogeneous external
electromagnetic field which has strong Fourier components for the
required momentum transfer. It has been suggested to use the strong
electric fields of a crystal lattice for this purpose~\cite{Bragg}. In
practice one can use germanium detectors which were originally built
to search for neutrinoless double-beta decay and for WIMP dark
matter. The Ge crystal serves simultaneously as a ``transition agent''
between solar axions and x-rays and as an x-ray detector. The beauty
of this method is that one can piggy-back on the existing Ge
experiments, provided one determines the absolute orientations of the
crystal axes relative to the Sun because the expected conversion rate
depends on the lattice orientation in analogy to Bragg diffraction.

A first limit produced by 
the SOLAX Collaboration~\cite{Solax} of
$g_{a\gamma}\alt30\times10^{-10}~{\rm GeV}^{-1}$ is not yet
self-consistent as the properties of the Sun already require
$g_{a\gamma}\alt10\times10^{-10}~{\rm GeV}^{-1}$.  However, the limit
easily cuts across the axion line (it applies for $m_a\alt 1~{\rm
keV}$), and no doubt it can be significantly improved as $\beta\beta$
and WIMP search experiments grow in size and exposure time.

\subsection{M\"ossbauer Absorption}

\vskip2pt

If axions essentially decouple from photons for $E/N=2$ models, and if
they also do not couple to electrons at tree level, there is a small
window of allowed axion masses in the neighborhood of $10~{\rm eV}$
(Fig.~\ref{fig:limits}). One can search for axions in this range by
appealing only to their coupling to nucleons. The Sun would emit a
nearly monochromatic $14.4~{\rm keV}$ axion line from thermal
transitions between the first excited and ground state of $^{57}$Fe
which is quite abundant in the Sun. In the laboratory one can then
search for the axion absorption process which would give rise to
x-rays as $^{57}$Fe de-excites~\cite{Moriyama}. Of course, the Doppler
broadening of the line in the Sun of about $5~{\rm eV}$ is much larger
than the natural line width of order $10~{\rm neV}$ so that the
M\"ossbauer absorber in the laboratory picks up only a small fraction
of the total flux. Even so it may be possible to detect or
significantly constrain solar axions in an experiment which is now in
preparation in Tokyo~\cite{Moriyama98}.  A recent pilot experiment by
another group did not have enough sensitivity to find axions in the
above window~\cite{Krcmar}.

%%%%%%%%%%%%%%%%%%%%%%%%%%%%%%%%%%%%%%%%%%%%%%%%%%%%%%%%%%%%%%%%%%%%%%
%% Section V %%%%%%%%%%%%%%%%%%%%%%%%%%%%%%%%%%%%%%%%%%%%%%%%%%%%%%%%%
%%%%%%%%%%%%%%%%%%%%%%%%%%%%%%%%%%%%%%%%%%%%%%%%%%%%%%%%%%%%%%%%%%%%%%

\section{SUMMARY}

A surprisingly large number of experiments to search for solar and
galactic dark-matter axions have recently emerged. The U.S.~Axion
Search as well as the Kyoto experiment CARRACK have now reached a
sensitivity where they could realistically detect galactic dark matter
axions, surely an important step because the role of axions as an
alternative to supersymmetric particles as a cold dark matter
candidate is perhaps the most important aspect of the continuing
interest in axion physics. As it stands, axion dark matter could well
show up before the millenium ends!

%%%%%%%%%%%%%%%%%%%%%%%%%%%%%%%%%%%%%%%%%%%%%%%%%%%%%%%%%%%%%%%%%%%%%%
%% Acknowledgments %%%%%%%%%%%%%%%%%%%%%%%%%%%%%%%%%%%%%%%%%%%%%%%%%%%
%%%%%%%%%%%%%%%%%%%%%%%%%%%%%%%%%%%%%%%%%%%%%%%%%%%%%%%%%%%%%%%%%%%%%%

\section*{ACKNOWLEDGMENTS}

Partial support by the Deutsche Forschungsgemeinschaft under grant
No.\ SFB-375 is acknowledged.

%%%%%%%%%%%%%%%%%%%%%%%%%%%%%%%%%%%%%%%%%%%%%%%%%%%%%%%%%%%%%%%%%%%%%%
%% References %%%%%%%%%%%%%%%%%%%%%%%%%%%%%%%%%%%%%%%%%%%%%%%%%%%%%%%%
%%%%%%%%%%%%%%%%%%%%%%%%%%%%%%%%%%%%%%%%%%%%%%%%%%%%%%%%%%%%%%%%%%%%%%

\end{document}